\documentclass[journal,twocolumn,letterpaper]{IEEEJERM}
\usepackage{ifpdf}

\usepackage{cite}

\ifCLASSINFOpdf
   \usepackage[pdftex]{graphicx}
\else
\fi
\usepackage{amsmath}
\usepackage{array}
\usepackage{fixltx2e}

\usepackage{stfloats}

\usepackage{times,amsmath,epsfig}
\usepackage{fancyhdr}
\usepackage{amsmath}
\usepackage{amsfonts}
\usepackage{amssymb}
\usepackage[latin1]{inputenc}
\usepackage{array}
\usepackage{graphicx}
\usepackage{url}
\usepackage{subfigure}
\usepackage{bm}
\usepackage{breqn}
\usepackage{xcolor}
\usepackage{soul}
\usepackage{amssymb}
\usepackage{comment}
\usepackage{tabularx}

\begin{document}

%
\title{AMICal Sat: A sparse RGB imager on board a 2U cubesat to study the aurora}

%
\author{Mathieu Barthelemy (1)(2),
                    \and
          Elisa Robert (1)(2)(6),
                    \and
        Vladimir Kalegaev (3),
       		\and
	Vincent Grennerat (2)(4),
		\and
	Thierry Sequies (2)(4),
		\and
	Guillaume Bourdarot (1)(5),
		\and
	Etienne Le Coarer (1)(2),
		\and
	Jean-Jacques Correia (1)
		\and
	Patrick Rabou (1)

        (1): Univ. Grenoble Alpes, CNRS, IPAG, F38000 Grenoble, France\\
        (2): Univ. Grenoble Alpes, CSUG, F38000, Grenoble, France\\ 
        (3): MSU-SINP, Leninskie gory, GSP-1, Moscow 119991, Russian Federation \\
        (4): Univ. Grenoble Alpes, IUT1, F38000 Grenoble, France\\
        (5): Univ. Grenoble Alpes, CNRS, LIPHY, 38000 Grenoble, France\\
        (6): SpaceAble, F75009 Paris, France
}

\markboth{IEEE The Journal on Miniaturization for Air and Space Systems}
{M. Barthelemy \MakeLowercase{\textit{et al.}}: AMICal Sat}

\twocolumn[
\begin{@twocolumnfalse}
  
\maketitle

\begin{abstract}
AMICal sat, a dedicated 2U cubesat, has been developed, in order to monitor the auroral emissions, with a dedicated imager. It aims to help to reconstruct the low energy electrons fluxes up to 30 keV in Earth auroral regions. It includes an imager entirely designed in Grenoble University Space Center. The imager uses a 1.3 Mpixels sparse RGB CMOS detector and a wide field objective (f=22.5 mm). The satellite platform has been built by the polish company Satrevolution. Launched September, $3^{rd}$, 2020 from Kuru (French Guyana) on board the Vega flight 16, it produces its first images in October 2020. The aim of this paper is to describe the design of the payload especially the optics and the proximity electronics, to describe the use of the payload for space weather purpose. A preliminary analysis of a first image showing the relevance of such an instrument for auroral monitoring is performed. This analysis allowed to reconstruct from one of the first images the local electron input flux at the top of the atmosphere during the exposure time. 
\end{abstract}

\begin{IEEEkeywords}
cubesat, imager, aurora.
\end{IEEEkeywords}

\end{@twocolumnfalse}]

%
\IEEEpeerreviewmaketitle

\section{Introduction}
%
%
%
%


Space weather is a system science in the sense that it includes a chain of complex phenomena coming from the Sun and going to the Earth mainly through the magnetosphere. Added to this, the effects on the Earth infrastructures and their vulnerability have to be taken into account. All this chain is too poorly described to allow accurate nowcasting and forescasting of the space weather events and of their effects on Earth \cite{Schrijver2015}. In this chain, the upper atmosphere as well as its interface with the magnetosphere require improvements in their description.

Precipitations of auroral electrons along magnetic lines lead to auroras, which are one of the most striking manifestations of space weather. These phenomena characterize the relationship of the magnetosphere and the upper atmosphere, and their intensity and localization indicate the state of near-Earth space. The energy release in the region of the auroral oval, associated with precipitation of auroral electrons, is controlled by the solar wind parameters and is one of the important reasons leading to changes in space weather in the polar magnetosphere and ionosphere.

It is therefore fundamental to observe, monitor and model the energetic inputs in the relevant regions of the near-Earth space. Koskinen et al. (2017)\cite{Koskinen2017} highlight the strong importance of improving the accuracy of both measurements and numerical models. In this frame, one of the main gaps in both data and modelling is the monitoring of the precipitation of low-energy $(0.02-30 keV)$ particles in the ionosphere and in the magnetosphere, especially electrons which are key contributors to ionospheric currents. 

Ground-based networks of instruments for auroral monitoring have been used for several decades. Among those currently active, we can mention the MIRACLE network of all-sky cameras, the ALIS network \cite{Kauristie2001}, \cite{Brandstrom2001} or the ASK \cite{Dahlgren2011} instruments which can produce very fast images in several emission bands. However, although very useful, ground-based instruments are often blinded by cloud coverage and cannot be operated with daylight. For instance, weather conditions in Ny-Alesund on Svalbard are cloudy about 75\% of the time \cite{Cisek2017}. These constrains hence drastically limit optical data availability for space weather applications.

Other ground-based instruments can provide complementary observations of upper-atmospheric processes and magnetosphere-ionosphere couplings. We can mention for example incoherent scatter radars like EISCAT or magnetometers networks on the ground. However, these data are not fully suitable to monitor the low-energy particles coming into the atmosphere in auroral regions or they cover limitated part of the oval preventing to get large scale monitoring. Optical data are thus mandatory. 
In this frame, space observations of auroral emissions can bring invaluable information, since they are not affected by cloud coverage, they allow to monitor wavelengths absorbed by the atmosphere, and the geometry allows limb observations of aurora, which makes it possible to reconstruct the vertical profile of the emissions, a task which is rather difficult from the ground and impossible with a single instrument.

\section{Space-born imager for auroral monitoring}

Numerous satellites observed the polar lights both in the UV and visible, however AMICal Sat is the first cubesat to be dedicated to the observation of the optical emissions of the auroras. 
Most of them are observing the auroras in the UV especially in the Far UV. For example, we can quote the following missions:
\begin{itemize}
\item IMAGE satellite with FUV (Far UltraViolet imaging) instrument \cite{Mende2000}, 
\item TIMED satellite with GUVI instrument \cite{Christensen2003}, 
\item Freja satellite with ultraviolet imager \cite{Lundin1995}, 
\item Dynamic explorer satellite with SAI (Spin Scan Auroral Imager) instrument \cite{Rees1988} 
\item Viking satellite with V5 UV instrument \cite{Anger1987}. 
\item Polar satellite with UVI (Ultraviolet Imager) instrument \cite{Germany1998}, 
\item Feng Yun satellite with WAI (Wide field Auroral Imager) instrument \cite{Zhang2019} with a very wide field $2\times10^{\circ}\times68^{\circ}$, 
\item DMSP satellite with SSULI (Special Sensor Ultraviolet Limb Imager) and SSUSI (Special Sensor Ultraviolet Spectrographic Imager) \cite{McCoy1992}, \cite{Paxton1992}.
\end{itemize}

 Much less satellite are observing auroras in the visible. We can mention:
 \begin{itemize}
 \item NPS (Nadir-looking Photometer System) on board DMSP
 \item The instrument VIS (Visible Imaging System) \cite{Frank1995} on board Polar Satellite. 
 \item Lastly, Cassiopee small-satellite carries near infra red \& narrow band visible imager with the FAI (Fast Auroral Imager) instrument \cite{Cogger2015}. 
 \end{itemize}
 
 However in the visible range the Japanese satellite REIMEI is one of the most interesting.

REIMEI is a small satellite operated by the Institute of Space and Astronautical Science, Japan Exploration Agency (ISAS/JAXA) with the aim to understand interaction with high atmosphere and auroral electrons. It has been launched August 23, 2005 from Baikonour by Dnepr rocket on a polar sunsynchronous orbit at 608 by 655 km of altitude and inclination of $98.17^{\circ}$. This small-satellite have a mass of 72 kg and size of 70 cm x 62 cm x 62 (H) cm. It carries four instruments : 
\begin{itemize}
\item A multi-spectral monochromatic auroral imaging camera (MAC),
\item an electrons and ion energy spectrum analyzers(ESA/ISA), 
\item a plasma current probes (CRM) and
\item a three-axis geomagnetic field aspect sensor (GAS).
\end{itemize}

GAS instrument allows the auroral camera MAC, to capture the projected footprint of REIMEI position along the local magnetic field line at altitude of 110 km. ESA/ISA instrument measures electrons and ions carried by the magnetic field line and responsible for the auroral emissions captured by MAC. REIMEI is then able to measure auroral particle properties and auroras images simultaneous, is called Mode-S.
MAC instrument has a field of view of ~70 x 70 km, in the local horizontal plane direction and a matrix of $1024 \times 1024$ pixels for three wavelenghts bands : 427.8 nm, 557.7 nm and 670 nm. Temporal resolution is 120 ms. 
ESA/ISA instrument allows to obtain energy-pitch angle distribution function of auroral particles with the energies of 10 eV/q - 12keV/q for ions and 12 eV - 12 keV for electrons \cite{Saito2011}\cite{Ebihara2010}. Intensities observed by MAC must be corrected as explain in \cite{Whiter2012} due to parasitic light sources. They mentioned Moon light pollution and reflection of the aurora itself on the ground or clouds.

One of the remaining challenges in the frame of auroral particle quantification, is to cover the full oval in order to reconstruct a larger part of the energetic inputs. Several way could allow to achieve this : First is to provide a very wide field instrument, the other is to build a constellation of satellites. To keep the cost at quite low levels the most convenient way for this last option is to use cubesats.

\section{AMICal Sat imager}

AMICal Sat, a 2U  demonstrator cubesat, has been launched September $3^{rd}$ on board the Vega flight 16. It is on a sun synchronous orbit with a mean altitude of 530 km with an inclination of $97.5^{\circ}$. The payload is an imager designed to take pictures of the aurora both at the limb and the nadir. It aims to allow short exposure time in the range of 1 second. The imager is equipped with an home made objective with a focal length of 22.5 mm and an aperture of f/1.4. The imager fits in less than 1U.

The real originality of this imager is the detector with its sparse RGB CMOS detector using $10 \mu m $ pixels. As stated by Sharif and Jung \cite{Sharif2019}, the black and white pixels are much more sensitive than colored ones. This can also been checked looking at the datasheet of the ONYX detector. Only one pixel over 16 is colored. This detector named ONYX contains 1.3 millions of pixels ($1280\times1024$).

Since the launch, AMICal Sat is working and has been able after commissioning to begin the science mission in October 2020. However a failure in the ADCS (Attitude Determination Control System) perturbs the mission since the orientation control is thus lost. However images has been registered. They show both the aurora and the ground on the night side.

\subsection{Auroral observations with sparse RGB detector}

Auroral emissions have the particularity to be an emission line spectra and not a continuum. Main lines in the visible in term of intensity are the $O^{1}S$ green line at 557 nm, the $O^{1}D$ triplet at 630 nm and the 0-1 band of the $N_{2}^{+}$ first negative system at 427 nm. In the very near UV it also exist the 0-0 band at 391 nm of the same electronic transition of this ion which show a stronger intensity than the 427 band with a constant branching ratio. It is not visible with the AMICal Sat instrument.

$N_{2}$ first positive bands also have significant intensity especially when considering the integration over large spectral bands especially between 600 nm and 700 nm. 

The fact that in the blue and green filter band pass, the intensity of these main lines represent the larger part of the intensity is interesting when using large band filters to monitor aurora. In a first approximation, the green line intensity will be equal to the intensity in the green pixels as well as the 427 line in the blue filter. The situation is immediately more complicated in the red filter bandpass, where several bands are present especially the first positive band of $N_{2}$ and the oxygen red line.

In black and whites pixels, the intensities of all lines will be mixed. However those pixels are much more sensitive. The global sensitivity will then be much higher since they represent 15/16 of the total number of pixels. This will be useful for the shape reconstruction.

However the sensitivity of the filters is not zero even far away from the central wavelength meaning that a slight component of the other lines is also present. If we consider only the lines mentioned above we can consider that each filter is a linear combination of the different lines. The coefficients are given in table \ref{Parameters}. 
\begin{table}
\centering
\caption{\bf Linear combination of each filter regarding main auroral lines}
\begin{tabular}{p{1.8cm}p{0.9cm}p{0.9cm}p{0.9cm}p{0.9cm}p{0.9cm}p{0.9cm}}
\hline
Line & Wavelength (nm) &  Blue Filter &  Green Filter & Red Filter & Panchromatic pixels \\
\hline
$O^{1}S$ & 557 &0.084 & 50.06 & 0.0418 & 0.6618 \\
$O^{1}D$ & 630-636-639 & 0.12 &0.085 & 0.663 & 0.663 \\
$N_{2} (A^{3}\Sigma_{u}^{+} - B^{3}\Pi_{g}) (\Delta v = 3) $ & 630-680 &0.14 &0.11 & 0.65& 0.653  \\
$N_{2} (A^{3}\Sigma_{u}^{+} - B^{3}\Pi_{g}) (\Delta v = 4) $ & 570-610 &0.06 &0.32 & 0.39\footnote{The Quantum efficiency (QE) of the red and green filters strongly varies at these wavelength. We considered an average} & 0.65  \\
$ N^{2}_{+} (B^{2}\Sigma_{u}^{+} - X^{2}\Sigma_{g}^{+}) (0-1)$ &427&0.425 &0.038 & 0.0735 & 0.5113 \\

\hline
\end{tabular}
  \label{Parameters}
\end{table}

In the AMICal Sat configuration each pixel represents 240 m at the ground level considering a mean altitude of the satellite of 530 km. Since the pattern for red and blue pixels is 8 pixels in each direction, this means that we obtain red information only every 2 km. This the same for the blue pixels. However the $B \& W$ pixel resolution stays close to 240 m.

It means that we keep the  information on the geometrical structure of the aurora despite the lack of colored information.

The reconstruction of the colored image can be done via a color reconstruction via a switch to YCbCr representation. However for science purpose it is better to distinguish both $B \and W$ images from colored channels. We then perform an interpolation of each color using cubic functions and obtain 4 different images, one $B \and W$ and three colored ones.

\subsection{Optical design}

The central part of the imager is a compact, fast, and wide field objective lens, suited for a cubesat platform. These challenging characteristics were not met in any existing commercial solutions and have called for a specific development in AMICal Sat mission.\\

The parameters achieved in the design of the imager are summarized in Table \ref{tab:Parameters2}. The choice of the parameters were driven as follows : first, a pixel size of $10\mu \mathrm{m}$ of the Onyx detector was chosen, which enables a compromise between a large optical \textit{\'etendue} (faint targets) and small spatial resolution. Then, the size of the detector and the total field of view ($40^\circ$) set the focal length of the design. From an initial f-number of f/1.1 (targeted), the requirement on the aperture was revised to f/1.4, once again to be compatible with short exposure of faint targets. Finally, the total optical length (entrance pupil to detector) had to be smaller than 50mm, in order to fit the 1U volume dedicated to the payload.

\begin{table}[h!]
\centering
\caption{\bf Design parameters of the optical payload}
\begin{tabular}{cc}
\hline
Parameter & Value \\
\hline
Spectral domain & 405-768nm \\
Pixel size & $10\mu\mathrm{m}$ \\
f-number & f/1.4  \\
Total field of view (Diagonal) & $40.8^\circ$  \\
Focal length & 22.5mm  \\
Total optical length & 42.9mm \\

\hline
\end{tabular}
  \label{tab:Parameters2}
\end{table}

\begin{figure}[!h]
\centering
\includegraphics[width=3.2in]{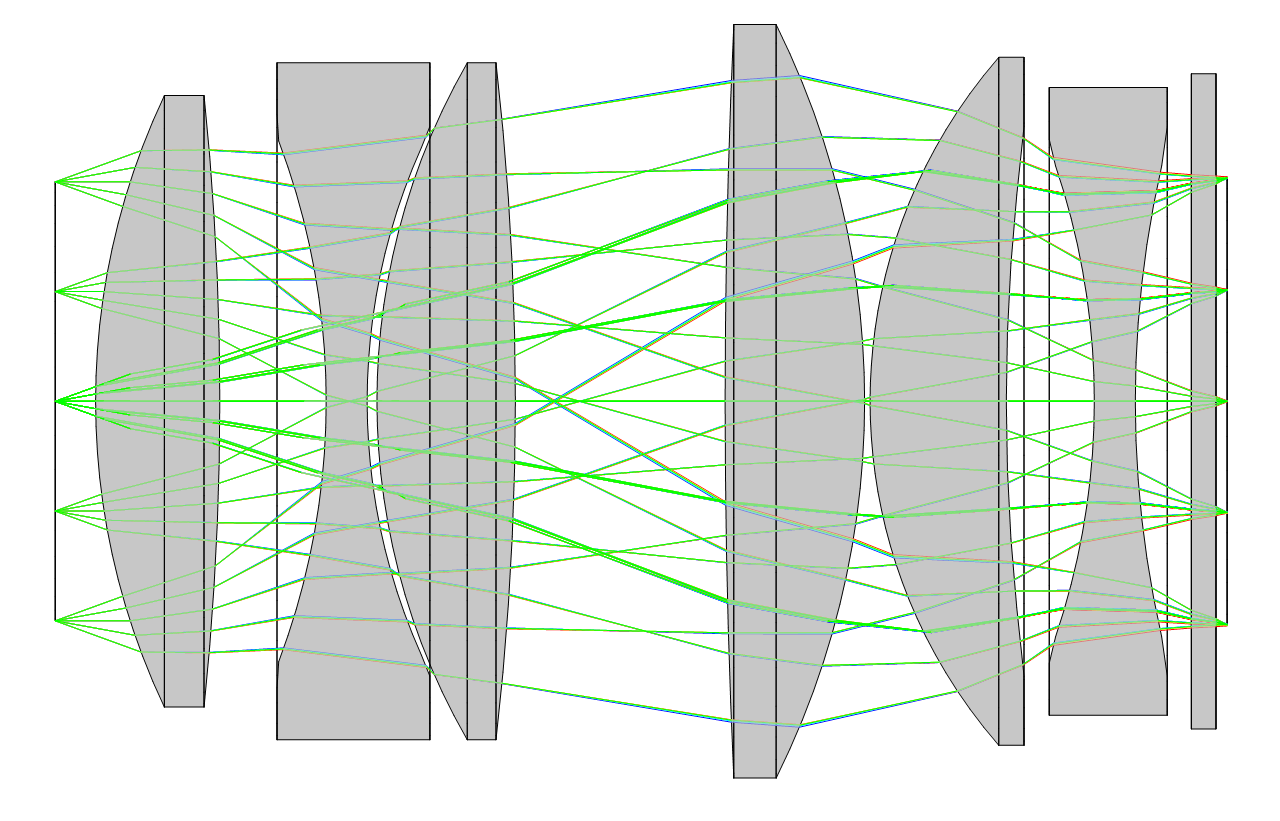}
\caption{\bf AMICal Sat optical design. Left pupil, Right detector}
\label{fig_ImagerDesign}
\end{figure}

The design adopted here is of the kind of double Gauss objectives (Fig.\ref{fig_ImagerDesign}), with a special emphasis put on the compacity of the design. The resulting solution includes 6 lenses, made of two glass materials (S-LAH65VS for lens 1, 3, 4 and 5; S-TIH14 for lens 2 and 6) in order to compensate for the chromatic dispersion, at the level of each air-spaced doublets. These doublets are intentionally not cemented, in order to be compatible with the void conditions in space.

\begin{figure}[!h]
\centering
\includegraphics[width=1.8in]{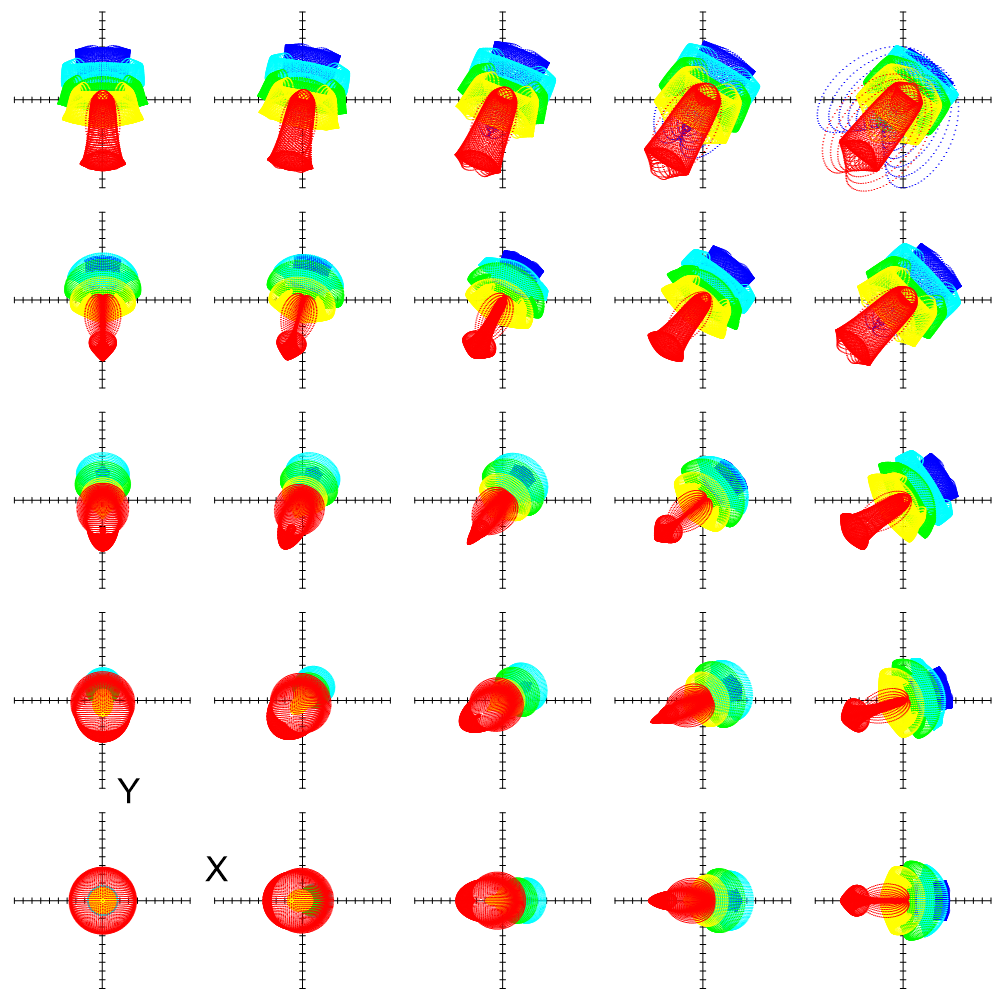}
\caption{\textbf{Imager spot diagram. Field of view quarter} (axis scale is $5\mu \mathrm{m}$)}
\label{fig_PSF}
\end{figure}

The typical size of the Point Spread Function (PSF), defined as the edge of the square that encircles 80\% of the energy of the PSF, is $14.96\mu \mathrm{m}$ at the center of the field, $33.6\mu\mathrm{m}$ at half-diagonal, and $47.52\mu \mathrm{m}$ at the extreme edge of the diagonal (Fig. \ref{fig_PSF}). The objective was also specifically designed to avoid vignetting. It is not rigorously telecentric, although an important constraint was put on the maximum incidence angle tolerated on each pixel, in a form of image telecentric design. This maximum incidence angle is set by the design of the pixel itself and its embedded microlens.

For the sake of both compacity and image quality, the last lens of the design acts as a field lens, and was placed at close separation of the optical window of the detector. These features (close back-focus, close air-spaced doublets) called for a specific mechanical development, which were essential to reach the performances of the instrument. \\

\subsection{Baffle and mechanical design}
\begin{figure}[!h]
\centering
\includegraphics[width=1.8in]{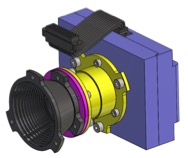}

\caption{\textbf{Global view of the mechanical design}}
\label{fig_mec_design}
\end{figure}

Amical-SAT imager can be separated in 3 parts, the optical baffle to remove the stray light, the optical imager and the detector with its electronic board (Fig. \ref{fig_mec_design}).

As stated above, 6 lenses are collecting the aurora light and focus it on the detector. There are positioned by spacers with an important accuracy (between 10 and 20 micrometers). The important thermal cycling between day time and night time, forced us to choice a material with same coefficient of thermal expansion than the lenses. Titanium alloy (Ta6V) is chosen for all the spacers.

To avoid important stress inside the lenses glass, lenses and spacers are fixed with a group of spring. The springs load is determined with the total weight of the lenses and spacers and the maximum acceleration during the launch phase. A margin factor of 1.5 is applied.
The distance between the lenses and the detector can be adjusted by a shim spacer. Spacers are machining with holes and grooves to remove air during the launch phase.
To reduce the weight, the mechanical box is made with aluminum alloy (7075 T6). The material of all the screws is stainless steel 316 with class 8.8. All the aluminum parts are coated with standard black anodized. The titanium spacers are not coated. Due to the baffle and the pupil stop, there is no direct light on those spacers.

\begin{figure}[!h]
\centering
\includegraphics[width=3.2in]{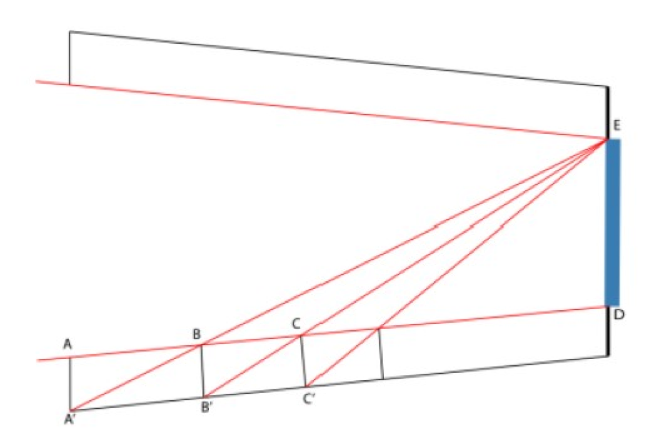}
\caption{\textbf{Baffle vane design method following \cite{Scaduto2006}}}
\label{fig_Vane}
\end{figure}

To avoid parasitic light out form field of view, a baffle has been designed. It considers the following constraints: 
\begin{itemize}
\item{Available space,}
\item{Field of view and optical properties, }
\item{Structural constraints to avoid vibration of the objective since the baffle links the optical mount to the satellite structure,}
\item{Thermal protection of sensor,}
\item{Manufacturing difficulties}. 
 \end{itemize}
To maximize the efficiency of the baffle, custom vanes have been designed. Fig. \ref{fig_Vane} described the ray tracing method for building the vanes as explained in \cite{Scaduto2006} or in \cite{Fest2013}. There are no coating on the mechanical parts, even on lenses spacers. 


The shape of vanes is difficult to machin. Grenoble Alpes University (S.MART resource platform) is exploiting a Electron Beam Modeling (EBM) equipment. This process is deposing metallic powder, under vacuum, which is particularly compatible with space environment. Added to this 3D printing allows to get high roughness. In this case, it was larger than $3.2 \mu m$ giving better optical properties.

\begin{figure}[!h]
\centering
\includegraphics[width=3.2in]{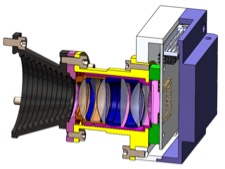}

\caption{\textbf{Imager opto-mechanical implementation}, including baffle and detector.}
\label{fig_design_tot}
\end{figure}

The global design is summarized in Fig. \ref{fig_design_tot}.

We had decided to delegate the machining of the optical imager parts and the integration at the same company (OPA-Opticad) than the lenses. The baffle and the electronic box was machined at Grenoble University facility. The final integration of the imager was done at IPAG clean room (ISO 5).

Before launch, tests were done only at acceptance level in the integrated nanosat. Random vibration tests were done by Air Liquid company (ALAT- France) and the thermal tests were done by Satrevolution SA (Poland).
The vibration random level requirement specified are in table. \ref{VibParameters}.

\begin{table}[h!]
\centering
\caption{\bf Vibration level for mechanical tests}
\begin{tabular}{ccc}
\hline
3 Axes vibrations & Level $(g^{2}/Hz)$ & Global ($G_{rms}$)\\
Frequency range & 1 min at 0 dB & \\ 
\hline
  
 20-50 Hz& 0.0071 &   \\
100 Hz& 0.0142& \\
200-500 Hz& 0.0355 & 5.89\\
1000 Hz & 0.0142 & \\
2000Hz & 0.0071 & \\
\hline
\end{tabular}
  \label{VibParameters}
\end{table}

The CubeSat withstands the random vibrations at acceptance level on X, Y and Z.

\begin{figure}[!h]
\centering
\includegraphics[width=3.2in]{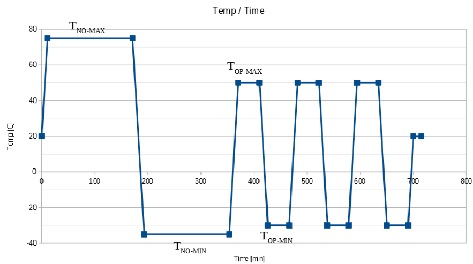}

\caption{\textbf{Thermal cycling of the payload.}}
\label{fig_thermal}
\end{figure}

Two different thermal tests were done, thermal bake-out and thermal cycling between $-35^{\circ}C$ and $75^{\circ}$ during 700 min (Fig. \ref{fig_thermal}). The CubeSat withstands at the both thermal test at acceptance level.

\subsection{Optics transmission}

At this stage, due to a failure in the ADCS of AMICal Sat, it has been impossible to take any picture of the moon. The sensitivity is thus the one calculated from ab-initio calculation using the specificity of the detector and optical front end elements. The simulation of the optical elements gives the transmission for the objective described in table \ref{tab:transmission}. The coating leads to a transmission between 0.98 in the green and 0.95 in the red. It has significant effect in the blue since the transmission in then only 0.8.

\begin{table}[h!]
\centering
\caption{\bf Transmission of the obectives vs wavelength}
\begin{tabular}{p{1.8cm}p{1cm}p{1cm}p{1cm}p{1cm}p{1cm}}
\hline
Lambda (nm) & 0.43583   &     0.48613    &    0.58756    &    0.65627   \\
\hline
Transmission lenses    &	0.9657  &  0.9854  &  0.9957  &  0.9949  \\
\hline
Transmission coating & 0.8 & 0.97 & 0.95 & 0.95 \\
\hline
Total transmission 	&	0.7725   &  0.9558  &   0.9459	& 0.9451  \\
\hline

\end{tabular}
  \label{tab:transmission}
\end{table}

We thus obtain a total transmission for the objective, varying from 0.77 to 0.96, then by multiplying by the quantum efficiency of each filter and each detector, we are able to calculate the sensitivity of the instrument. 

 \subsection{Line intensity reconstruction} 

We express the intensity in total flux for each pixel for each extracted images considering that each pixel has a field of view of $2.05\times10^{-7} sr$

The full equation to reconstruct the intensity emitted in auroral region is then for one line:

$I_{g}=S_{pix} \times \Omega_{pix} \times (QE_{g 557} \times Abs_{g} \times L_{557} +\\
 QE_{g 427} \times Abs_{b} \times L_{427} +\\
  QE_{g 630} \times Abs_{r} \times L_{630} + \\
 QE_{g N2 \Delta v=3} \times Abs_{r} \times L_{N2 \Delta v=3} + \\
 QE_{g N2 \Delta v=4} \times Abs_{r} \times L_{N2 \Delta v=4})$\\

$I_{r}=S_{pix} \times \Omega_{pix}  \times (QE_{r 557} \times Abs_{g} \times L_{557} + \\
QE_{r 427} \times Abs_{b} \times L_{427} +\\
 QE_{r 630} \times Abs_{r} \times L_{630} + \\
QE_{r N2 \Delta v=3} \times Abs_{r} \times L_{N2 \Delta v=3} + \\
QE_{r N2 \Delta v=4} \times Abs_{r} \times L_{N2 \Delta v=4})$\\

$I_{b}=S_{pix} \times \Omega_{pix}   \times (QE_{b 557} \times Abs_{g} \times L_{557} + \\
QE_{b 427} \times Abs_{b} \times L_{427} +\\
 QE_{b 630} \times Abs_{r} \times L_{630} + \\
QE_{b N2 \Delta v=3} \times Abs_{r} \times L_{N2 \Delta v=3} + \\
QE_{b N2 \Delta v=4} \times Abs_{r} \times L_{N2 \Delta v=4})$\\

$I_{Pan}=S_{pix} \times \Omega_{pix}   \times (QE_{Pan 557} \times Abs_{g} \times L_{557} + \\
QE_{Pan 427} \times Abs_{b} \times L_{427} +\\
 QE_{Pan 630} \times Abs_{r} \times L_{630} + \\
QE_{Pan N2 \Delta v=3} \times Abs_{r} \times L_{N2 \Delta v=3} + \\
QE_{Pan N2 \Delta v=4} \times Abs_{r} \times L_{N2 \Delta v=4})$\\

where QE is the quantum efficiency at a given wavelength, Abs the absorption due to the optical elements, and L the received luminance for a given line. S represents the surface of the pixel and $\Omega$ the solid angle for one pixel.

As it is, the system is underdetermined since it contains 5 variables for only four equations. We can however consider a constant branching ratio between the $\Delta=3$ branch and the $\Delta=4$ of the first positive band. $\frac{I(\Delta v=4)}{I(\Delta v=3)}=0.21$, reducing the number of variables. Based on laboratory spectra from \cite{Mangina2011}, the $\Delta v=3$ and $\Delta v=4$ represent more than 95\% of the total intensity of $N_{2}$ first positive band. We can then consider that these two bands represents the full first positive band and we choose not to correct and consider the full intensity is included in these spectral features.
Added to this we considered the uncertainties due to the noise in the detector as described in the ONYX datasheet. It indicates at $25^{\circ}C$ :
\begin{itemize}
\item{Readout noise ERS: $\leq 3$}
\item{DSNU: $10$}
\item{PRNU: $\leq 1$}
\item{Dark signal: $2$}
\end{itemize}
However the temperature of the satellite is included between $0^{\circ}C$ and $-10^{\circ}C$, much lower than $25^{\circ}C$. We then can consider that the noises are much lower probably less than 5 counts. To evaluate the effect of this noise we propagate a noise of 5 counts coded in 12 bits in the processing pipe. Considering the bias described above and the noises described in the datasheet, the intensity reconstruction precision is then evaluate of the order of 10\% for low signal, down to 2\% for almost saturated signals. 
However some variable lines not taken into account in the previous calculations could also degrade this precision. Extremely difficult to estimate since they are variable but never higher than few percent, we then consider that we have 10\% uncertainties for high signal also.

\subsection{Line of sight}

After intensity reconstruction, it is necessary, in order to compare with auroral emission intensity, to reconstruct the line of sight.

The luminance obtained on the instrument is :
\\

$L = \int_{0}^{z_{min}} Em_{vol} . \Omega R^{2} \frac{dz}{cos(\theta)}$
\\

where $\theta$ is the angle between the line of sight and the local vertical and $Em_{vol}$ is the volume emission rate in the aurora which can be compared with auroral simulation outputs.

\section{Electronics design}

\subsection{Institutional framework}\label{subsec_Inst}
The institutional framework in which the design and fabrication of the electronics took place, led to some non-optimal technical solutions. It has to be briefly exposed in order to give a better understanding of these.

AMICal Sat payload electronics specification started about 20 months before the delivery date of the payload. Although the electronics team
was directed by a single person throughout the whole project duration, several students were successively involved, mostly for a duration of 2 to 6 months. This turnover combined with the short duration of the project led us to prioritize choices such as:
\begin{itemize}
\item use of technologies already executed in labworks of the university, thus well known by staff and students,
\item easier and faster prototyping solutions, immediately doable at the university labs,
\item buying from common suppliers, hence limiting the possibilities of using space qualified parts.
\end{itemize}
ONYX sensor part was an exception, because it came to the electronics team as an input of the project. Therefore, next subsections also detail what the team was able to set up, in order to eventually design and build a fully functional payload, despite the team limited knowledge and lack of experience and support on this part.
\subsection{Inputs of the project}
Nano-satellites common design constraints often include:
\begin{itemize}
\item limited volume allowed to the payload electronics,
\item low power design and advanced sleeping modes.
\end{itemize}
In AMICal Sat case, the choice of allocating 1 full U (1~dm$^ 3$) to the payload, combined with the size of the optics and its sensor, led to some confortable room allocated to the PCB (Printed Circuit Board). In the end, it is rather high frequencies considerations (differential serial outputs of the sensor, and fast Static RAM signals) that conducted us to miniaturize the PCB. Indeed, the power supply design wasn't room constrained at all, and no tradeoff had to be made between PCB room and quality of the power supplies decoupling (capacity and max voltage values). Regarding the power consumption, the ONYX sensor rules a minimum working frequency, which defined the whole payload acquisition working frequency. When implementing sleeping modes on embedded systems, a balance has to be done between:
\begin{itemize}
\item energy cost of the wake-up from sleep versus full boot-up and reconfiguration from powerOff state,
\item On/Idle time ratio,
\item waking up periodicity.
\end{itemize}
In our case, the balance was clearly in favor of a full powerOff of the payload between 2 sets of acquisitions.
\newline

Other inputs that had to be dealt with were:
\begin{itemize}
\item high throughput of the ONYX sensor (total throughput cannot be set lower than 975~Mbits/s),
\item low noise analog power supplies to the sensor,
\item 12 bits DCMI (Digital Camera Media Interface) bus to the SatRevolution OBC, with a throughput limited to about 60~Mbits/s because of OBC limitations,
\item when possible, space qualified design.
\end{itemize}

\begin{figure}[!t]
\centering
\includegraphics[width=3.5in]{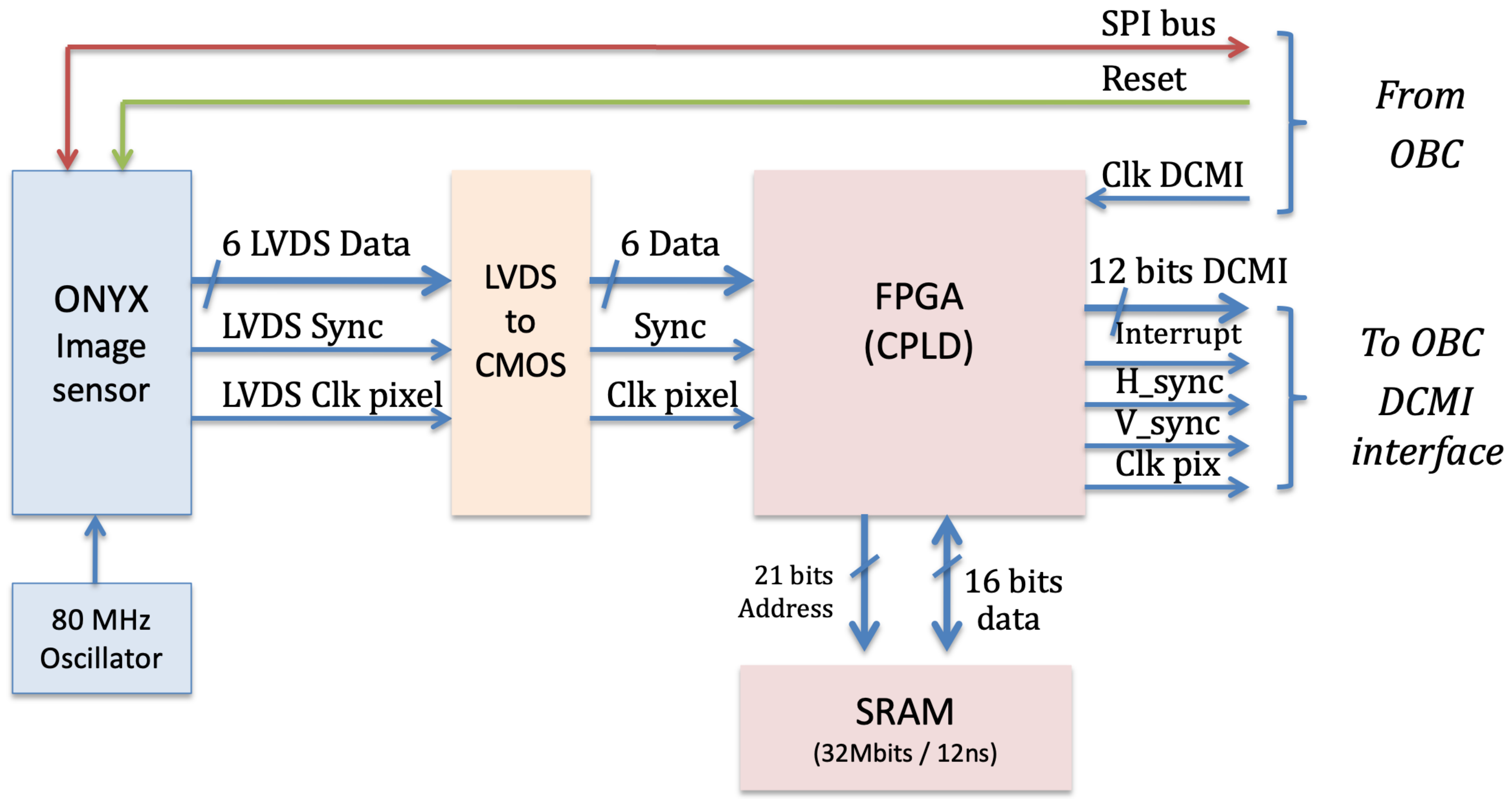}
\caption{\bf AMICal Sat payload Digital dataflow}
\label{fig_DigitalDataFlow}
\end{figure}

\subsection{Digital design in a space qualified context}
Basically, the payload electronics is an image frame grabber. Because of the great gap between sensor minimum throughput and OBC 12 bits DCMI input bus throughput (limited to 7~Mwords/s), the digital architecture detailed in Fig. \ref{fig_DigitalDataFlow} was chosen:
\begin{itemize}
\item a small FPGA (actually a big CPLD) to de-serialize the sensor's 6 serial outputs,
\item a 32 Mbits SRAM to bufferize a full frame.
\end{itemize}
When starting the digital design, we studied the ESCC QPL (European Space Components Coordination Qualified Parts List). We immediately noticed how very few complex digital components are qualified. This led us to consider using non-space-qualified parts, but with extra care during the parts selection.
Space qualified FPGAs (Field Programmable Gate Array) were eliminated from the choice list because of their overpriced IDE tools and the lack of experience of staff and students on it. The Altera mature MAX II CPLD (Complex Programmable Logic Device) family was chosen because of its compliance with the institutional framework described in \S \ref{subsec_Inst}, and also because of its availability in automotive-grade. One of the big concerns about FPGAs / CPLDs in space, is the risk of alteration of their configuration memory. Anti-fuse memory technology is well suited, but as already said, no such part entered in our institutional framework. MAX II family is based on EEPROM configuration memory, which is much better than more recent Flash technology in terms of TID (Total Ionizing Dose) and SEE (Single Event Effect) tolerance \cite{MemTech:Troxel}.

The frame buffer memory has to be at least 12 bits wide, in order to write each 12 bits pixel in a single write cycle, faster than 12~ns. These constraints shrunk the choice to a couple of easily available SRAM parts. SRAM has better TID and SEE tolerance than DRAM  (Dynamic RAM) \cite{MemTech:Troxel}, but it was chosen over DRAM mainly because of its much simpler implementation of the device access. MRAM (Magnetic RAM) memory, a technology used on AMICal Sat platform as the OBC main memory, has the best TID and SEE tolerance \cite{JPL:MRAM}. But in 2017, the only STT-MRAM available technology, did not meet the 12~ns write cycle requirement.

As previously explained, the SRAM and MAX II CPLD TID and SEE tolerances are rather good. On top of that, choice was made to shield the whole payload PCB electronics in a 3~mm thick aluminum case, itself mounted inside the 1~mm aluminum external walls of the nano-satellite. CSUG ran aluminum thickness simulations with the OMERE tool, for AMICal Sat orbit. Results show that for a 1 to 3 years mission, the TID is lowered by a factor 12 when using a 4 mm Al shielding compared to a 1 mm Al shielding. These particules will therefore mainly hit the ONYX Sensor. Since then, some radiation testing on recent FPGA devices were made, showing that some FPGAs and their aside PROM memory can be a good choice for space tolerance, and thus should not need such shielding \cite{Xilinx:TID-SEE}.

ONYX sensor is space hardened by CSUG sponsor and provider Teledyne-E2V. On our design, accordingly to Teledyne-E2V, we chose to electrically protect it against SEL (Single Event Latch-up). This point is detailed in the next section.

DCMI bus, control lines, and power lines (with redundancy) represent a total of 39 lines to reliably connect platform OBC to payload. Space qualified connectors and link were provided by CSUG's sponsor Nicomatic.

To conclude on the digital design, we insist on the fact that we were led to choose some non-space-qualified parts. But the careful selection of digital components, based on their TID / SEE tolerant technology and automotive grading, combined with the use of the metal shielding, assured us to have the best chances of success.

\subsection{Payload power supplies design}
\subsubsection{Single Event Latch-up protections}
\begin{figure}[!t]
\centering
\includegraphics[width=3.5in]{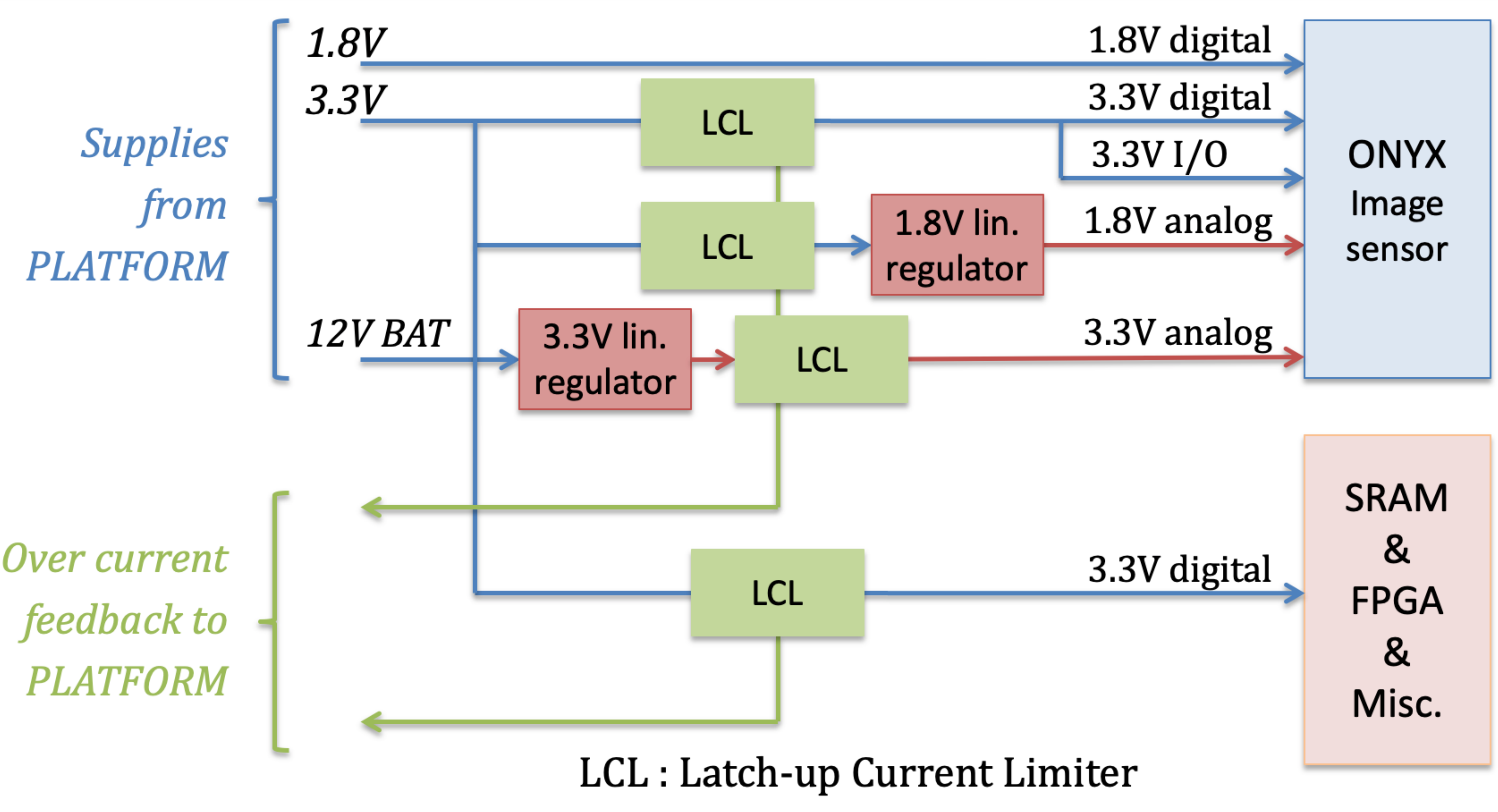}
\caption{\bf AMICal Sat payload power scheme}
\label{fig_PowerScheme}
\end{figure}

platform provides the payload with 1.8~V, 3.3~V and unregulated 10 to 13~V from batteries. 1.8 and 3.3 lines are generated by DC-DC switching power supplies with residual, uncharacterized switching noise. 3.3 line is well suited to directly power the digital parts of the payload. Therefore, payload power design was mainly about ONYX sensor analog power supplies and SEL protection.
In order to achieve a high level of SEL protection, it is important to tune precisely the over-current triggers, for each supply line. And for return of experience, it is interesting to get separate feedbacks from sensor SELs and from digital electronics SELs. Thus, four current limiting switches (LCL: Latch-up Current Limiter) were implemented (see on Fig. \ref{fig_PowerScheme}).

Selected TI TPS2553-1 part acts as circuit breaker, with a 2~$\mu$s over-current response. Trigging levels were set to 2 to 3 times the maximum drawn current, based on datasheets informations. Lab tests and characterizations confirmed these values. The LCL TPS2553-1 has a 6.5~V input limit. Therefore, on the 3.3V analog line, it is placed on the downstream side of the voltage regulator. To preserve the analog line accuracy, we compensate the small voltage drop across the LCL, thanks to a 3.3V linear regulator with feedback sense input connected to the output of the LCL.
\subsubsection{ONYX Sensor power supplies}
in order to reach ONYX theoretical pixel readout low noise, analog power lines are regulated with low noise (less than 30$\mu$Vrms) linear regulators. These are a significant source of power waste aboard. During a typical active state of 1~s per picture, 1.8V analog line draws 36~mW from the 3.3V line thus wasting 30 mW in heat. But 3.3V analog line draws 148~mW from the 12V line, resulting in a big waste of 390~mW (to be compared to the 937~mW global consumption of the ONYX sensor). This only 69\% efficiency is the consequence of a tradeoff avoiding the addition of a 12V-5V DC-DC converter (Fig. \ref{fig_PowerScheme}).

The ONYX sensor power-up sequencing between the several analog and digital lines is done through discrete electronics (RC circuit + transistor). This avoids using a PMIC (Power Management Integrated Circuit) complex IC, subject to qualification considerations.

\subsection{Space qualification considerations}
Although constrained by our institutional framework, efforts were made to carry design and fabrication closer to a space qualified product.
\begin{itemize}
\item Active parts were chosen among automotive AEC-Q200 certified parts when available, or at least in industrial temperature range version.
\item Passive parts are easily available in AEC-Q200, and were all chosen accordingly.
\item Capacitors voltage rating was systematically doubled.
\item The main risk for ceramic capacitors is a mechanical break of their structure, due to launch vibrations. This usually leads to fatal short circuit of the corresponding power line. Cascaded electrodes technologies like the AVX Flexisafe product line lowers drastically this risk. We were not able to provision such parts, and had to settle for more standard AEC-Q200 parts.
\item A trick to limit the latter risk, consists in splitting the decoupling capacitor in two equal capacitors, serial mounted, with the drawback of doubling the room occupied on PCB. When we realized we could not get hold of cascaded electrodes capacitors, it was too late to modify the PCB.
\end{itemize}
The flying model PCB was manufactured in N35 space qualified material with ENEPIG plating. Prototyping and flying model PCB assembly was conducted at CEDMS, a UGA university lab, part of the IUT1 Electrical Engineering department:
\begin{itemize}
\item soldering material was standard SnPb solder,
\item soldering process of the SRAM small pitch BGA package was elaborated through several tests, thanks to the great experience of the lab technician\footnote{Pierina Pierotti, UGA, IUT1 Dpt. GEII},
\item in order to absolutely avoid the risk of a weak intermetallic soldering, any form of hand soldering was ruled out, except for the PGA ONYX package,
\item critical soldering zones like BGA inter-balls spaces were visually checked with dedicated optical instruments, and full X-ray imaging checking was conducted.
\end{itemize} 

\subsection{Sharing of operation experience}
At the time of writing, analysis of the first datasets we got from operating  AMICal Sat shows that:
\begin{itemize}
\item payload electronics is successful at powering and controlling the ONYX image sensor,
\item payload digital design is successful at grabbing, buffering, and pre-processing a frame,
\item lack of some random pixels in images is not related to payload electronics, but to S-band communication considerations,
\item payload single event latch-up feedback signals are ineffective, that matter is analyzed below.
\end{itemize} 
We can thus conclude that the overall precautions taken during the design of AMICal Sat electronics were adequate to reach as close as possible to a space qualified design, despite our institutional context constraints.

After the launch and first payload power-up, examining the telemetry informations immediately showed evidence of a minor error in the electronics design: IC TPS2553-1 fault output signal is open-collector, active low. Resistors pulling-up to the 3.3V line supply were implemented on the payload PCB. When the platform OBC powers down the payload, this 3.3V line becomes floating, and so does the fault signal, generating spurious faults to the OBC. Pull-up resistors should have been implemented on the OBC PCB, above the 3.3V payload line switch. Also, onboard OBC software is inefficient at memorizing pertinent SEL faults which could happen during payload ON state: it needed to be processed as an interrupt, with timestamp storage of the event. This part of the software, which could not be provided directly within the payload API (Application Programer Interface) to SatRevolution OBC sub-contractor, should have been subject to a specific software interface document, instead of just an oral recommandation. Since this minor malfunction could have been partly corrected during the final tests after assembly and integration, it demonstrates the importance of very comprehensive test specifications, even on non-critical features.

Another interesting experience feedback is the CPLD choice. It proved at first to be technically pertinent during the prototyping phases A \& B of the project, but began to show poor digital electronics behavior in late phase C, when testing the design's full performance. A misinterpretation of the ONYX sensor documentation (not yet stabilized at that time), about how the deserialisation should be carried out in the CPLD or FPGA, led to instability in the frame grabbing process. At that point, switching to a more appropriate FPGA, would have required a large redesign of the PCB. It would have unacceptably delayed the payload electronics delivery. Thankfully, this problem was smartly circumvented in the last months, by tricking the HDL design and performing extensive signal stability testing. Today, that very difficult choice of standing for a design on which the team already had strong skills, proved to be successful.

\section{Images}

Despite the problem in the ADCS, it has been possible to take some images. Several of them can be exploitable for auroral monitoring. We will focus on one of them quoted in the following pictures 46R\footnote{NB: R and L refer to the SD card where the the images were registered} (Table \ref{tab:images}).

\subsection{Timestamp}
During the commissioning phase of the satellite mission, we experienced a shift in the master clock of the satellite. This shift has been measured and quantified. We thus give as timestamp for each picture the corrected one.

\begin{table}[h!]
\centering
\caption{\bf Images characteristics.  }
\begin{tabular}{p{0.7cm}p{1.05cm}p{1.05cm}p{1cm}p{0.8cm}p{0.6cm}p{0.6cm}}
\hline
Pic number & Timestamp   & Corrected timesptamp  & Date & Time (UT) & Lat & Long \\
\hline
 & 160335+4 digits & 160335+4 digits & & & & \\
 \hline
46R & 6795 & 6808 & 22/10/2020 & 8:53:28 & $73.46^{\circ}$ S & $129.80^{\circ}$ W  \\

\hline
\end{tabular}
  \label{tab:images}
\end{table}

\subsection{Ground based reflection}
Two sources of light can perturb significantly the auroral emission measurement. First one is the moon, second one is the aurora itself both reflected on the clouds or snow cover. Estimation of the moon intensity can be done since the moon is a very stable photometric standard and a quasi punctual source. 

The intensity of the Moon depends of the phase of the Moon. As a reference, we can consider the following calculations have been done based on \cite{Cramer2013} and considering the following hypothesis:
\begin{itemize}
\item Date of the data: 30 November 2012, 11:40:43. Almost full moon, $96\%$
\item Flux at the top of the atmosphere: $2.633 \mu W.m^{-2}. nm^{-1}$ at 550 nm (Uncertainty $0.45\%$)
\item Cloud/snow albedo: 0.7, No specular reflexion, reflexion in $2\pi sr$ (Lambertian)
\item Nadir sighting. 
\end{itemize}
 
 The intensity on each filter is then:
\\

 $L_{Moon Filter}=\int_{\lambda_{min}}^{\lambda_{max}} F_{top} QE(\lambda) \times A \times \frac{\lambda}{hc} \times \frac{1}{2\pi} d\lambda$
\\

 where A is the albedo and the other parameters are the usual ones. The integral of the moon spectra between 400 nm and 700 nm represents 290.2 kR most of the time, much larger than the auroral emission. 
 At the time and location of picture 46R snap, the moon elevation was $17.65^{\circ}$ at the ground altitude and the phase was 37\%. We then can count on a small moon contribution in the image if clouds or snow are present. We neglected them in the following calculations. In further developments, the ROLO and POLO codes to calculate the moon intensity at the top of the atmosphere will be used. However hypothesis will have to be done on the snow and cloud coverage except if coordinated measurements in the infrared are not available. Since our calculation are based on the hypothesis of the absence of cloud or snow, we also neglect the reflexion of the aurora itself.

\section{Auroral interpretation}
The interpretation of the auroral intensities are made using the Transsolo code \cite{Vialatte2017}, a kinetic code which use as input the electron flux and the solar EUV flux on the dayside. It calculates the transport of the suprathermal electrons along a line of sight or a vertical and the subsequent auroral emissions. Emission output is the volume emission rate at different altitude as mentioned in section Line of Sight.  A schematic view of the Transsolo flow is given in \cite{Barthelemy2018}. 

\subsection{Nadir pointing}
\begin{figure}[!t]
\centering
\includegraphics[width=2.8in]{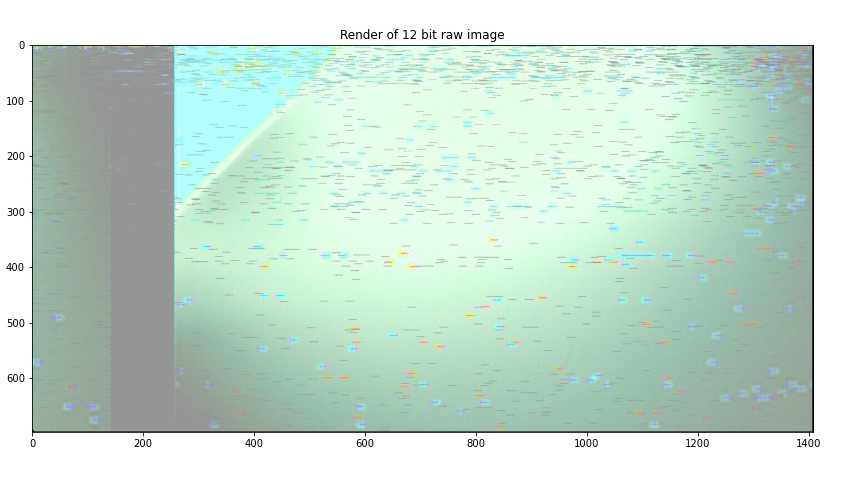}
\caption{\bf{Color reconstructed image taken by AMICal Sat. Axes represent pixels numbers. Some image packets have been lost during radio transmission and give dispersed color dots.}}
\label{image_46}
\end{figure}

Image 46R pointed the terrestrial globe. Some parts of the image are close to direct nadir pointing. Fig. \ref{image_46} shows the color reconstructed image. Intense colored points are lack in the radio transmission. The left band should not be positioned like this but should be repositioned on the right. The black band is a reference.

As mentioned in section Optical Design, the PSF varies along the diagonal from $14.96\mu \mathrm{m}$ at the center of the field to $47.52\mu \mathrm{m}$ at the extreme edge. To avoid dispersion of the light energy on the edge of the field, we choose to integrate the intensity on a much larger square field. Since the objective has been designed to fully avoid the vignetting, we then consider that the sensitivity is constant over the entire field.
The sensitivity determination could have been much better using moon calibration as planned. However since the ACDS of the satellite is out of order, this method represents the only way to get absolute intensity.

Arbitrary we first chose to consider a $17 \times 17$ pixels region and reconstruct the input parameters i.e. mean energy of the distribution and total flux  at the top of the atmosphere. The hypothesis of a maxwellian energy distribution is done. However, the unknown geometry is a difficulty in this frame, especially when the line of sight is strongly inclined. In this case, it crosses a wide panel of regions and it is then necessary to sample the line of sight and to run several times the Transsolo code. 

We chose in the paper to keep only line of sight close to the vertical with only one needed run of Transsolo.
Further investigations will be driven to take into account the specific geometries accessible on each image. The topic here is to show in one example the principles of the electron flux reconstruction from the images.

Using picture 46R around pixel (1216, 644) which is close to nadir pointing, we found an intensity in the four channels equal to 2620 counts in the panchromatic channel, 2000  counts in the red channel, 1540 counts in the green one and 1390 counts in the blue one. The reconstruction of the green and blue lines gives $I_{G}=1.92 kR$, $I_{427}=2.30 kR$. 
The reconstruction of the bands in the red O 630 nm and $N_{2}$ first positive band are too imprecise due to the fact that the panchromatic and the red channels show similar QE in this region. We cannot be confident in their extracted values. We thus only keep the green and 427 nm lines. Under the hypothesis of no parasitic light at the nadir meaning no cloud or snow, we found by minimizing the differences between measured intensities and simulated ones, an electron distribution supposed  to be maxwellian, with a mean energy of $3.02 keV$ and a total flux of $5.17  erg.cm^{-2}.s^{-1}$. In case of high albedo, the total flux will be biased, however the mean energy mainly based on $I_{G}/I_{427}$ ratio will be almost unchanged.

If comparing with other auroral intensity simulation especially the GLOW code \cite{Solomon2017}, we can see that Transsolo needs higher energy flux to get an equivalent green line intensity. In our case, the difference is around 50\%. However the codes are somehow different since GLOW is a two stream model and Transsolo a multi-stream model (8 in our calculations). On the other side, the chemistry simulation included in GLOW is more elaborate than in Transsolo. As stated in \cite{Gronoff2012} for the Martian case, the uncertainty propagations of cross sections especially, in kinetic codes like Transsolo, can drive to large uncertainties in the final results. Further investigations must be driven to explain these discrepancies.
 
\section{Conclusion}
Since its launch AMICal have been able to take some pictures of the aurora despite the failure of the ADCS. First rough interpretations of the data have been performed. They however remain uncertain due to several problems:

The first one and most important is the uncertainty on the geometry of the line of sight due to the failure of the ADCS.

The second one is the lack of different filters which necessitates to make some hypothesis on the intensity of the $N_{2}$ first positive vibrational bands. It shows the importance to continue to develop spectrometers and spectro-imager.

The third one is the absence of re-calibration on the moon which could give better photometric precision than the ab-initio calculation performed in this paper.

The fourth is the calculation of the background intensities due to the moon and the reflexion of the aurora itself.

It is possible to reduce some of these bias by precise calculation of background intensities, but measuring the aurora in the visible with such  sparse RGB detectors needs hypothesis on the branching ratio on the $N_{2}$ emissions. More precise estimations need spectra and/or of course spectral images. The future ATISE \cite{Barthelemy2018} and Wide Field Auroral Imager (WFAI)\cite{LeCoarer2021} will allow to get the additional information.

However we showed that it is possible with a sparse RGB detector to get the intensity of the green and $N_{2}$  $427 nm$ auroral lines and then reconstruct roughly the electron input fluxes under calculation hypothesis. If the mean energy reconstruction is done with a relative confidence, the total flux is very uncertain as shown by using comparisons with other simulations like GLOW. As this stage, we are only able to get an order of magnitude with an uncertainty estimated around 50\%.  The extraction of the N2 and Oxygen red line is still more difficult and in most of the cases impossible. 
Since it allows rough estimation of the electrons fluxes but are very low cost a large monitoring of the aurora with such imagers could be considered for further space missions. Considering they fit in less than 1U, it demonstrates the interest of cubesats like AMICal Sat for auroral space weather monitoring. However, the best way to reconstruct these electrons fluxes is clearly to perform spectral measurements and best spectral imaging measurements like the Wide Field Auroral Imager instrument mentioned in \cite{Lecoarer2021}.

The development of AMICal also allowed to test in real conditions a non space-borne CMOS detector, the Teledyne E2V Onyx, with a specific electronics described here, and to design a very compact wide field imager and its very robust packaging. As a last conclusion, it is also important to mention that more than 50 students worked on this project from Bachelor to Master degree, as well as two PhD students.

\section{acknowledgements}
 This work has been funded by CSUG industrial partners via sponsorship (Air Liquide Advanced Technologies, Teledyne E2V, Nicomatic, Lynred, NPC system, Gorgy Timing). It has been partially supported by the Labex FOCUS ANR-11-LABX-0013. The AMICal Sat program has also been partly funded by PNST, by the SHM thematic program of CNES and by the Universit\'e Grenoble Alpes. The PhD work of Elisa Robert has been funded by SpaceAble. We thank the entire radio amateur community for their help in downloading and decoding the telemetry and the data especially Daniel Estevez (EA4GPZ) and Julien Nicolas (F4HVX). We also thank Sergei Krasnopolev from NILAKT.

%

\begin{IEEEbiographynophoto}{Mathieu Barthelemy}
Mathieu Barthelemy is professor at Grenoble Alpes University (IPAG-CSUG). Specialised in space weather especially in auroral monitoring, he is the PI of AMICal Sat. He also is the director of the Grenoble University Space Center since its creation in 2015.
\end{IEEEbiographynophoto}

\begin{IEEEbiographynophoto}{Elisa Robert}
Elisa Robert is a PhD student at IPAG-CSUG, she also works for the SpaceAble company as space weather scientist.
\end{IEEEbiographynophoto}

\begin{IEEEbiographynophoto}{Valdimir Kalegaev}
Vladimir Kalegaev is head of the laboratory of space research working in Skobeltsyn Institute of Nuclear Physics of Moscow State University. His area of research is the study of global electrodynamics of the Earth's magnetosphere: solar wind-magnetosphere coupling, dynamics of large-scale magnetospheric current systems, particle fluxes and magnetospheric magnetic field variations during magnetic storm. Vladimir Kalegaev is the head of Space Monitoring Data Center of SINP MSU and leads a scientific group that studies the radiation conditions in the near-Earth space environment.
\end{IEEEbiographynophoto}

\begin{IEEEbiographynophoto}{Etienne Le Coarer}
Etienne le Coarer is engineer at  Grenoble Alpes University (IPAG-CSUG). Specialised in spectrometric  instrumentation for Astrophysics and planetology.
he introduced some new miniaturized spectrometry technics for small satellites. In AMICal Sat, he helps to size instrumentation for best performances.
\end{IEEEbiographynophoto}

\begin{IEEEbiographynophoto}{Vincent Grennerat}
Vincent Grennerat is instructor at Grenoble Alpes University (IUT1-GEII). His main teaching specialties are digital design, real time systems, and electronics. He has been involved with CSUG since 2015, mainly through student projects advising, and engineering of AMICal Sat.
\end{IEEEbiographynophoto}
\begin{IEEEbiographynophoto}{Thierry Sequies}
Thierry Sequies is engineer in mechanics at Grenoble Alpes University (IUT1-Mechanical engineering department). He has polyvalent skills, from design, manufacturing and iunderstand computering and architectures of hardware. He has been involved with CSUG since 2015, as mechanical supervisor and program manager for development of different space missions. 
\end{IEEEbiographynophoto}

\begin{IEEEbiographynophoto}{Jean Jacques Correia}
Jean-Jacques Correia is mechanical engineer at IPAG-CNRS. Specialised in mechanical and thermal design for space and telescope experiments.
\end{IEEEbiographynophoto}

\begin{IEEEbiographynophoto}{Guillaume Boudarot}
Guillaume Bourdarot is a PhD student in Astrophysics at Grenoble Alpes University (IPAG-LIPhy). His main expertise covers instrumental research for high-angular resolution in astronomy, and optical design of miniaturized payloads for space missions.
\end{IEEEbiographynophoto}


\begin{IEEEbiographynophoto}{Patrick Rabou}
Patrick RABOU is optical engineer at IPAG-CNRS, specialised in optical design for space and ground-based astronomical instruments. 
\end{IEEEbiographynophoto}




\end{document}